# A HYBRID SOFTWARE TEST AUTOMATION FOR EDUCATIONAL PORTALS

Aniwange, A.
Department of Computer Science & Engineering, Obafemi Awolowo University Ile-Ife
terteseamos@gmail.com

Nyishar, P. T.
Department of Computer Science & Engineering, Obafemi Awolowo University Ile-Ife
paulnyishar@gmail.com;

Afolabi, B. S.
Department of Computer Science & Engineering, Obafemi Awolowo University Ile-Ife
afolabib@gmail.com

Ejidokun, A. O.
Department of Computer Science & Engineering, Obafemi Awolowo University Ile-Ife
adebisiadetunji@gmail.com

**ABSTRACT**
Educational portal (EP) is a multi-function website that allows access to activities such as public and private sections, data retrieval and submission, personalized content and so on for the educational system. This study investigated the specific requirement for the enhancement of quality and behavior of EP with regards to time and cost using Obafemi Awolowo University (OAU), Ile-Ife, Nigeria as a case study. A test automation framework was designed using unified modelling language and implemented in Java programming language. MySQL and Excel database were used to store test data. The framework developed was evaluated using Test Time Performance (TTP), Performance Test Efficiency (PTE) and Automation Scripting Productivity (ASP) metrics. The results from the evaluation of the sample data provided showed that ASP produced a tested outcome of 360 operations per hour, PTE yielded 80% and TTP was just 4%. Based on the recorded performance, it is evident that the research can provide quick and firsthand information to quality assurance analyst and software testers, thereby reducing maintenance cost during software development.

## 1. INTRODUCTION
Software testing (ST) is an important part of the software development process (SDP) that is used to ensure software quality and behavior. Depending on the complexity and criticality of the product, it accounts for 30 to 60% of the life cycle cost [1]. It entails the developer's examination of individual modules (also known as unit testing), customer validation (acceptance testing), and monitoring of a network-centric service-oriented application's run-time [2].
Software testing can be done manually or with the assistance of software automation technologies. However, the complexity, pervasiveness, and criticality of software products are gradually increasing. As a result, it's critical to guarantee that software meets the essential standards of quality, consistency, and dependability. Using the manual method, this is quite tough to achieve [2]. Test automation (TA) is highly recommended to overcome the limitations of human testing since complex applications may be reviewed with less stress, in a short amount of time, and with a high degree of accuracy [3]. However, TA cannot operate in isolation; [4] believe that test automation requires a system for implementing basic capabilities such as test execution, monitoring, and reporting. This system must be scalable in order to develop a new form of test case known as Test Automation Frameworks and achieve reusability (TAF).
TAF is a work platform or support for automated testing that is defined by a collection of assumptions, concepts, and behaviors. It is in charge of designing the format in which expectations should be expressed, developing a mechanism to hook into or drive the application under test, running the tests, and reporting the findings. TAF is application agnostic, making it simple to scale, manage, and perpetuate. Given the complexity and relevance of EP applications, a TAF that best supports the automation testing of these applications should be considered.

## 2. RELATED WORK
[5] created a large-scale framework for automated test execution and reporting using the Python programming language, which merged data-driven and keyword-driven testing approaches. The results showed that the overall framework concept is possible when data-driven and keyword-driven testing are seamlessly integrated. [3] created a unique software test automation framework for complicated business processes. The framework took a keyword-driven approach, with test cases and data maintained in an excel spreadsheet. The outcome demonstrates that the planned framework is flexible and reusable across a variety of applications. Furthermore, compared to manual testing, it yields better results.
[6] presented a library architecture testing framework that allows code to be reused in methods and functions. Researchers reported, however, that it performed better than the Modular Based Testing Framework. [7] suggested an Android mobile data-driven architecture that uses an external database as input to test scripts that may be run several times for diverse inputs. [8] created an automated testing framework based on a data-driven approach and including the Selenium Automation Framework, which was written in Java (SAF). Nonetheless, the outcome shows that the framework is efficient and capable of handling the test case's vast size. However, a non-technical user cannot utilize the framework because the reusability of an application cannot be guaranteed.
[9] used a high-level object-oriented programming language to create a hybrid test automation framework for testing a web-based application (Java). Information was gathered by observation, survey, and interview procedures using an inductive or qualitative data analysis methodology. The findings revealed that automation testing is chosen over manual testing due to its reliability, ease of use, time savings, and reduced testing effort. [10] used a combination of library and data-driven architecture to build and implement a hybrid test automation framework. Because it takes advantage of different architectures, the framework is simple to use, adaptable, and maintainable.
The current automated test framework's weakness is that alternative reporting tools were not addressed or analyzed correctly, making it impossible to determine the robustness and efficiency of electronic portal applications. Furthermore, the literature framework was too broad and did not consider a specific case study like EPs. The goal of this research is to create and assess an integrated automated software test framework for educational portals.

## 3. METHODOLOGY
Data was collected using the qualitative research approach by the use of questionnaire, interviews and observations. Fifteen (15) respondents were interviewed using Google Hangout, WhatsApp, Stack and Trello via chat while ten (10) were engaged in a face-to-face interview using an open-ended questionnaire. The goal of the questionnaire was to inquire on the appropriate tool that can enhance the effective usage and easy deployment of the proposed framework in order to make it user friendly for flexible software tester and users not skilled in software development. The information extracted from the collected data to inform us on the choice of integration tools to be used for designing and implementing the framework.
The interactive framework design was done using UML diagrams: Class diagram and Activity diagram with the aid of Draw.io Unified Modelling Tool. The framework is very interactive in nature. It houses a data repository consisting of sets of reusable implemented methods specifically for portal applications. Also, it provides the flexibility for future integration of addition automation methods that may not be considered.
The class diagram described the structural overview and functionality of the framework implemented using object-oriented programming. The diagrammatic representation of the relationship between the classes is shown in the in Figure 1. It contains majorly 8 classes which are the testconfig, keywordsconstants, keyword, excelreader, mysqlreader, testing, report and keyword interface. Testconfig class contains the configuration variables for portal application under test while the keywordsconstants contains keyword constant for implementation. The keyword class implements the first two classes and it is responsible for test data processing. Excelreader provides a data repository (excel database) to the framework while mysqlreader provides interaction with the MySQL database. Testng class helps with writing of automated script for testing and report class generates report in automatic test report in html format. The keywordinterface class helps to drive test automation scripting.

Activity diagram describes the operational step-by-step workflows of components in the framework. It shows the overall flow of control, detailing the sequence of activities from a start point to the finish point, displaying the many decision paths that exist in the progression of events contained in the activity. Activity diagram for the framework is shown below in Figure 2.

## 4 FRAMEWORK IMPLEMENTATION

The framework was implemented using Java programming language. The developed framework supports test automation of any portal application that is powered by web technology. Different browsers can be used to test a portal application via the framework by integration of various web drivers. A Hybrid Software Test Automation Framework (HSTAF) for testing educational portal application as shown in Figure 3.

These techniques involve getting test data from a data storage location. The framework was able to test the login functionality of the portal automatically by reading the individual test cases from the data source and executing the same on the portal application. In this process, test data are read from either excel sheet or MySQL database based on the choice of the test engineer. The ExcelReader and the MysqlReader class are carefully implemented for this purpose. The ExcelReader makes use of the Apache poi library to read data from the excel files. On the other hand, the MysqlReader uses MySQL connector to connect with the MySQL database for the purpose of reading the test data. The second process is the automatic testing of the portal applications. In this process, the Keywords class is implemented to drive and power the automation testing. This class implements the TestConfig and KeywordConstant interfaces.

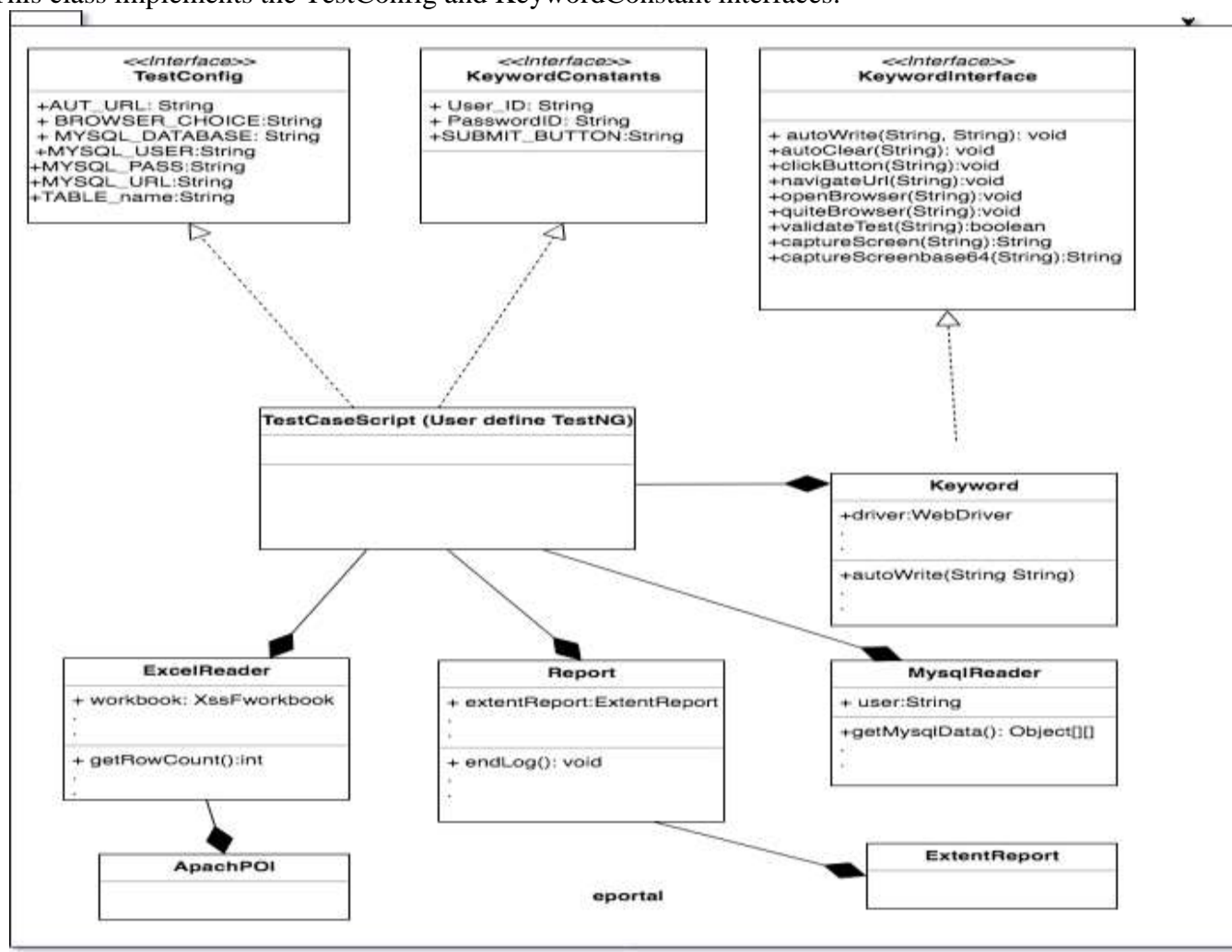

Figure 1: Class diagram of the test automation framework

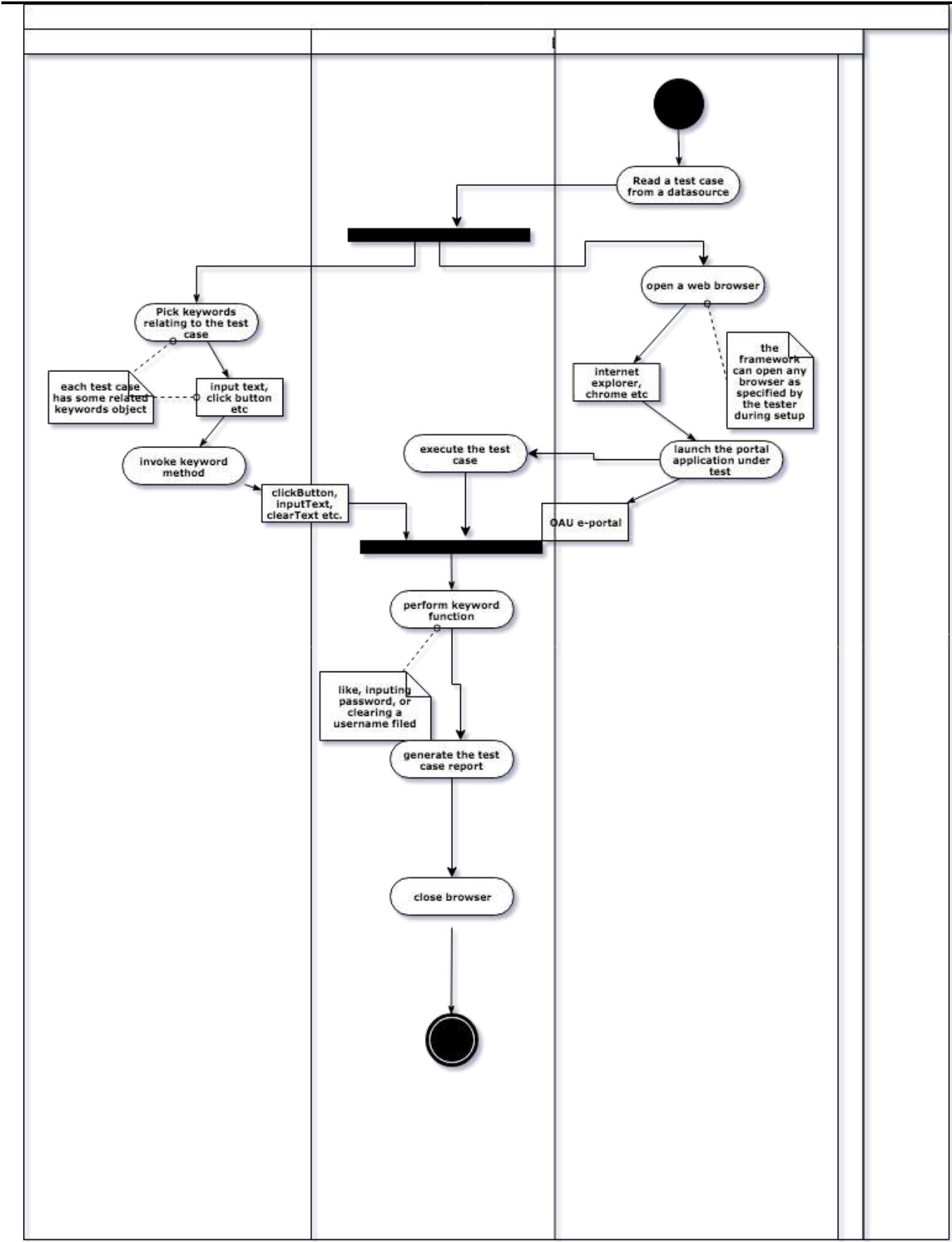

Figure 2: Class diagram of the test automation framework

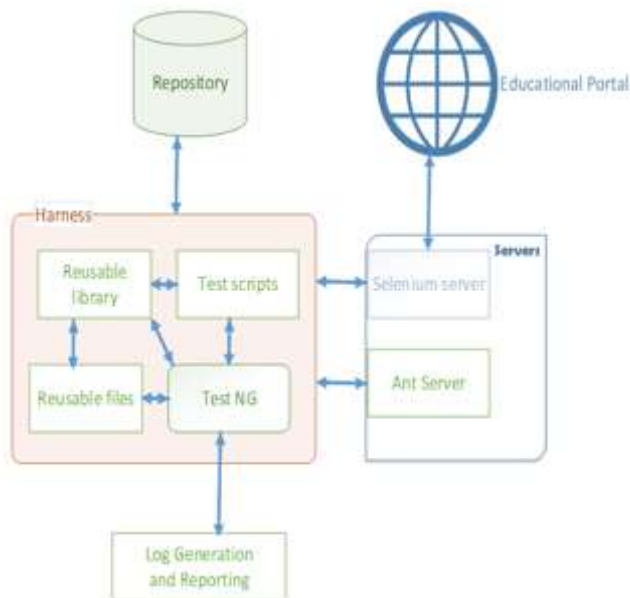

Figure 3: Developed hybrid automatic testing framework

Similarly, the Keyword class takes in the object of WebDriver, ExcelReader and MysqlReader, ExtentsReport and creates and an object that has the potential of encapsulating all the necessary methods (operations) that are needed to perform automation testing on the portal application. The last process of this framework is the generation of the test report. In this process, the ExtentsReport object is instantiated to generate both logs and the general report. Both reports are presented in a single HTML file with screenshot captures at the various stages of the test execution process. The framework generates an unambiguous test report to further support the analysis of the automation test cases. Figure 4 show how the Login test script was written in the framework as it makes use of the all the dependent methods to achieve the automation process.

The framework utilized the Report class of the framework that follows the standard report generation for automatic generation of test report, as is shown in Figure 5. The report contains a dashboard that summarizes the test execution process: the total number of steps, total test cases, and the percentage of passed and failed test cases, and total time of test execution. This report is clearly presented by the dashboard report of the framework. Furthermore, it contains information about the test environment, the user of the system, the type of OS, the Java version, and the Hostname of the system. There is navigation to individual test case detail report also in which information on the status (passed, failed, skipped, fatal, and so on) of each test case can be accessed. The framework was evaluated using test performance metrics which include: Test Time Performance (TTP) metric, Automation Scripting Productivity (ASP) metric and Performance Test Efficiency (PTE) metric.

Figure 4: The Login Test Script

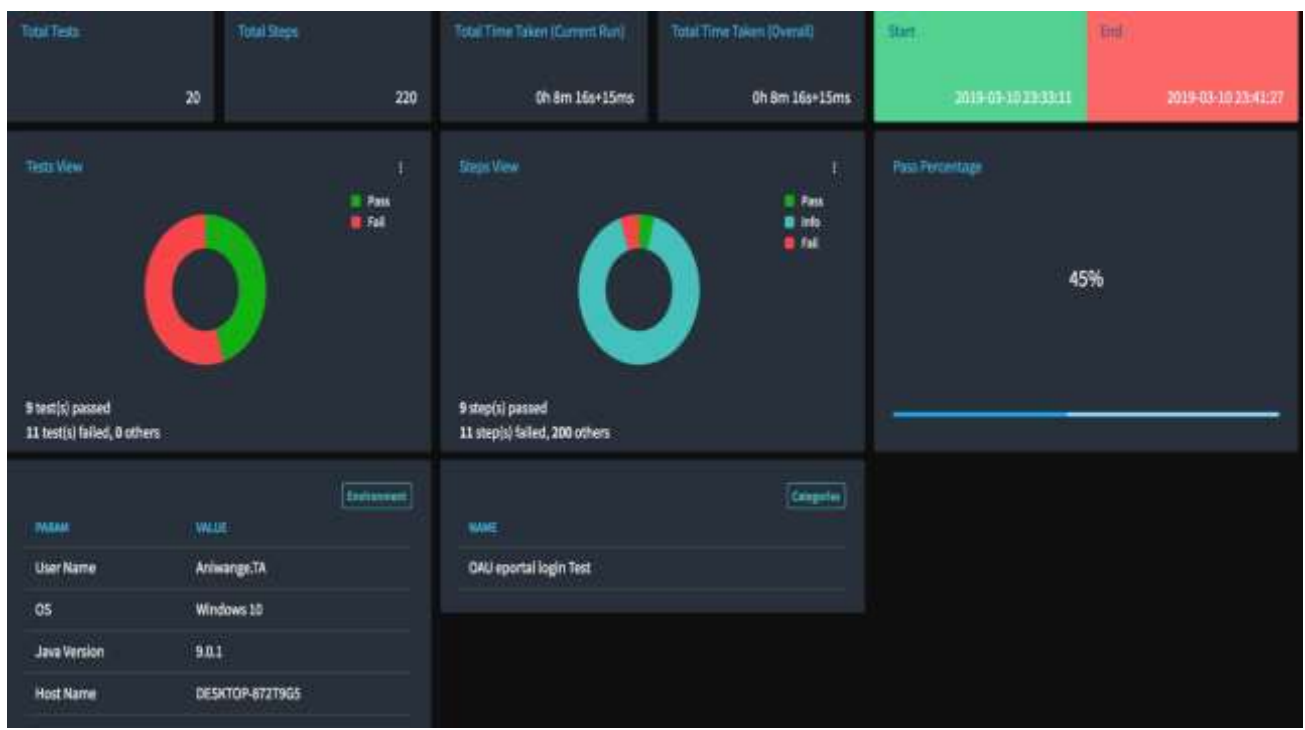

Figure 5: The Login Test Script

## 5 RESULT AND DISCUSSION
The educational portal test automation framework was evaluated based on the test time performance, performance test efficiency and automation scripting productivity. These metrics are used to assess the performance, effectiveness and quality of the framework.

### 5.1 Automation scripting productivity
From the test report generated by the test automation framework shown in Figure 4 and 5. Evaluation data were extracted to evaluate the framework using the performance metrics (ASP, PTE and TTP) the data extracted were presented in Table 1. The total operation performed was 180 in 30 minutes. The total number of operations per hour is 360 which is the Automation Scripting Productivity. This implies that with little or no training, an automation test script can easily be written to test the portal application function using the developed framework.

Table 1: Evaluation data table for ASP

| S/N | Operation Performed | Total |
|---|---|---|
| 1 | Number of click buttons | 20 |
| 2 | Number of browsers opened | 20 |
| 3 | Number of input parameters | 40 |
| 4 | Number of browsers closed | 20 |
| 5 | Number of validation performed on test cases | 20 |
| 6 | Number of the screenshot taken | 20 |
| 7 | Number of input field cleared | 40 |
|   | Total operation performed | 180 |
|   | The effort taken for scripting | 30 minutes |

### 5.2 Performance test efficiency
This metric determines the quality of the Performance testing team in meeting the requirements which can be used as an input for further improvisation if required. PTE can be calculates using the equation below:

$$Performance\ Test\ Efficiency = \left[\frac{\delta}{\delta + \varphi}\right] * 100\%$$

Where $\delta$ is requirement met before performance testing
$\varphi$ is requirement not met performance testing.

Table 2: Evaluation data table for PTE

| S/N | Requirement State | Number of requirements |
|-----|-------------------|------------------------|
| 1   | Before Test       | 4                      |
| 2   | After Test        | 1                      |
|     | Total             | 5                      |

The result of PTE gave 80%, which means the performance of the framework in meeting the automation requirement is relatively high.

**5.3 Test time performance**
This metric determines how fast it will take to test an application using the framework. The higher the percentage of these metrics, the more time is required to test the application. Similarly, the lower the metric, the less time required to test the application.
Figure 6 shows a bar chart of number of steps, the time, and the total number of test cases of the automation test execution. On the Framework performance, it was observed that 220 steps were performed during the test execution process and the total time taken to complete these steps was just 9 minutes. This gives an idea of how fast it is to perform an automation test on the framework

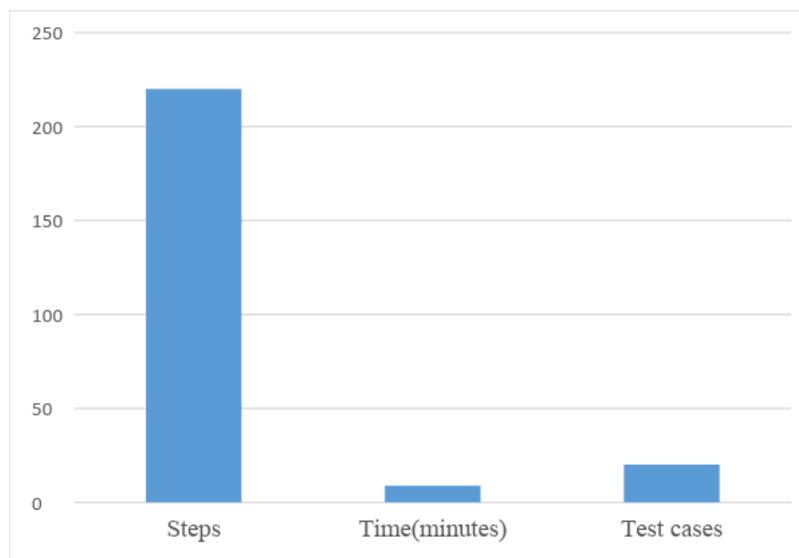

Figure 6: Automation test execution time analysis

Also, Test Time Performance (TTP) was just 4% test time efficiency. This is shown in Table 3. Consequently, it will take a small amount of time to test an entire portal application via this framework, therefore reducing the time spent on functional testing during portal applications development. It is arguably one of the quickest tools to test a portal application

Table 3: Evaluation summary calculations

| S/N | Metrics | Value |
|-----|---------|-------|
| 1   | PTE     | (4/ (4+1)) *100= 80% |
| 2   | ASP     | (180/0.5) = 360 Operations per Hours |
| 3   | TTP     | (9/220) *100 = 4% Minutes per Steps |

# 6 CONCLUSION

In this study, a hybrid software test automation framework for Educational Portal application was developed using recommended standards. The data acquired from the survey carried out, using questioner responses to ascertain the required tools for effective evaluation of educational portals. The Framework was evaluated using performance metrics such as PTE, ASP and TTP. The flexibility offered by the architectural design of this framework is a huge step forward in ensuring code reusability and its maintenance, which was a major drawback in automation framework implementation, over the last few decades. The research will not only reduce the number of test engineers in the development process. It also reduced the cost and time spent on manual testing of educational portals. Also, it serves as a base for future research in the area of test automation on educational portals. Subsequently, a framework that formulates good test cases for a portal application will be developed. The portal code optimization framework via code refactoring will be addressed to increase reusability and maintenance.